\documentclass[acmtog, nonacm]{acmart}

\usepackage{booktabs} % For formal tables
\usepackage[capitalize,nameinlink]{cleveref}
\usepackage{soul}
\usepackage{xfrac}
\usepackage{xcolor}

\usepackage[ruled]{algorithm2e} % For algorithms

\SetAlFnt{\small}
\SetAlCapFnt{\small}
\SetAlCapNameFnt{\small}
\SetAlCapHSkip{0pt}

\renewcommand\footnotetextcopyrightpermission[1]{}

\newcommand{\new}[1]{\textcolor{black}{#1}} %{red}{#1}}

\newcommand{\dataset}[1]{\textsc{#1}}
\newcommand{\ourrealds}{Two-Level Test}
\newcommand{\materialisticds}{Materialistic Test}
\title{Fine-Grained Spatially Varying Material Selection in Images}

\author{Julia Guerrero-Viu}
\affiliation{%
    \institution{Universidad de Zaragoza - I3A}
    \country{Spain}
}
\affiliation{%
    \institution{Adobe Research}
    \country{United Kingdom}
}
\email{juliagviu@unizar.es}
\orcid{0000-0002-2077-683X}

\author{Michael Fischer}
\affiliation{%
    \institution{Adobe Research}
    \country{United Kingdom}
}
\email{mifischer@adobe.com}
\orcid{0000-0002-2610-4831}

\author{Iliyan Georgiev}
\affiliation{%
    \institution{Adobe Research}
    \country{United Kingdom}
}
\email{igeorgiev@adobe.com}
\orcid{0000-0002-9655-2138}

\author{Elena Garces}
\affiliation{%
    \institution{Adobe Research}
    \country{France}
}
\email{elenag@adobe.com}
\orcid{}

\author{Diego Gutierrez}
\affiliation{%
    \institution{Universidad de Zaragoza - I3A}
    \country{Spain}
}
\email{diegog@unizar.es}
\orcid{0000-0002-7503-7022}

\author{Belen Masia}
\affiliation{%
    \institution{Universidad de Zaragoza - I3A}
    \country{Spain}
}
\email{bmasia@unizar.es}
\orcid{0000-0003-0060-7278}

\author{Valentin Deschaintre}
\affiliation{%
    \institution{Adobe Research}
    \country{United Kingdom}
}
\email{deschain@adobe.com}
\orcid{0000-0002-6219-3747}

\begin{document}

%%%%%%%%%%%%%%%%%%%%%%%%%%%%%%%%%%%%%%%%%%%%%%%%%%%%%%%%%%%%
% Teaser figure
%%%%%%%%%%%%%%%%%%%%%%%%%%%%%%%%%%%%%%%%%%%%%%%%%%%%%%%%%%%%

\begin{teaserfigure}
    \centering
    \includegraphics{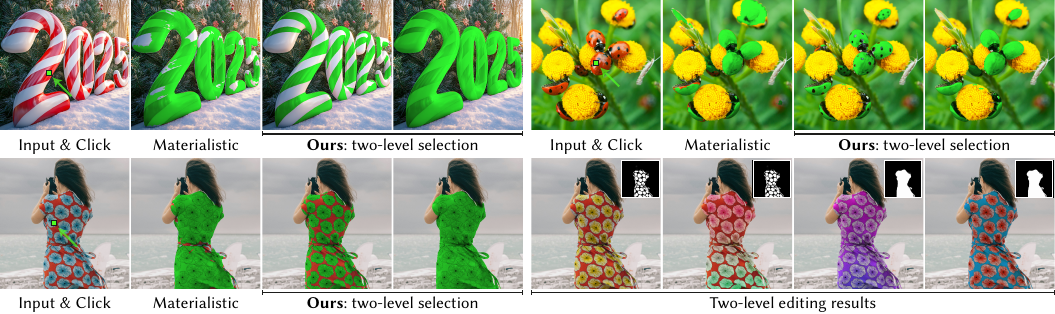}
    \vspace{-0.3cm}
    \caption{
        Our proposed method allows fine-grained material selection in images on two different levels of granularity, significantly outperforming previous work (Materialistic~\cite{sharma2023materialistic}) in selection accuracy and consistency. 
        We show here results on challenging examples due to specular reflections (top left) and fine patterns outside the training data (top right, bottom left).
        Selection masks are shown as green image overlays.
        \new{The bottom right row shows material editing results using our predicted two-level selection masks, with the masks shown as insets.} 
    }
    \label{fig:teaser}%
\end{teaserfigure}

%%%%%%%%%%%%%%%%%%%%%%%%%%%%%%%%%%%%%%%%%%%%%%%%%%%%%%%%%%%%
% Abstract
%%%%%%%%%%%%%%%%%%%%%%%%%%%%%%%%%%%%%%%%%%%%%%%%%%%%%%%%%%%%

\begin{abstract}
Selection is the first step in many image editing processes, enabling faster and simpler modifications of all pixels sharing a common modality. In this work, we present a method for material selection in images, robust to lighting and reflectance variations, which can be used for downstream editing tasks. We rely on vision transformer (ViT) models and leverage their features for selection, proposing a multi-resolution processing strategy that yields finer and more stable selection results than prior methods. Furthermore, we enable selection at two levels: texture and subtexture, leveraging a new two-level material selection (DuMaS) dataset which includes dense annotations for over 800,000 synthetic images, both on the texture and subtexture levels. 
\end{abstract}

\maketitle

\renewcommand{\shortauthors}{J. Guerrero-Viu et al.}

%%%%%%%%%%%%%%%%%%%%%%%%%%%%%%%%%%%%%%%%%%%%%%%%%%%%%%%%%%%%
\section{Introduction}
%%%%%%%%%%%%%%%%%%%%%%%%%%%%%%%%%%%%%%%%%%%%%%%%%%%%%%%%%%%%

Selection in images is an ubiquitous operation, enabling numerous downstream editing tasks. Given an image and a user input, such as a clicked pixel, selection aims to identify other pixels in the image that share a particular property with the user input.
Various modalities of selection exist, for example by color~\cite{belongie1998color}, object~\cite{ravi2024sam}, or material~\cite{sharma2023materialistic}.
The latter, in particular, facilitates selecting parts of an object or multiple objects easily, allowing further editing of regions that share the same material. 
Understanding which materials are the same in images also provides key information for inverse rendering and scene understanding tasks~\cite{nimierdavid2021material}. 

Recent work~\cite{sharma2023materialistic} has proposed to extract features from a pre-trained vision transformer (ViT)~\cite{caron2021dino} and spatially process them for selection. 
However, the ViT's tokenization patch size and small operating resolution limit the selection precision, especially around edges and thin structures.
Moreover, to fine-tune the ViT for material selection, \citeauthor{sharma2023materialistic} used a synthetic dataset which defined materials as \textit{textured} surfaces (e.g., a wallpaper with a repeating pattern texture); this  does not allow selecting image elements sharing similar \textit{subtexture} properties (e.g., a part of the repeating pattern with a similar appearance).

In this work, we tackle these limitations and propose a model for more precise and robust selection, capable of selecting at both texture and individual subtexture levels.
To extract more discriminative features for material selection we leverage DINOv2~\cite{oquab2024dinov2}. Its refined architecture and use of larger patch sizes to extract better contextual information outperforms other existing models like DINOv1~\cite{caron2021dino} or Hiera~\cite{ryali2023hiera}.
Unfortunately, the ViT's native resolution (518x518 for DINOv2) typically requires downscaling the input images before extracting features. This translates into poor performance when selecting boundaries or areas with fine details. To mitigate this, we devise a multi-resolution approach, splitting the original input image into tiles, and processing both a downscaled version of the original image and the tiles at the native ViT resolution, before concatenating them. 
This allows our aggregated features to capture both the high-level context provided by the downscaled image and finer-grained details from the better-resolved individual tiles, significantly enhancing selection quality. 

We also enhance the training process by sampling multiple pixels from various materials in each training image, instead of a single pixel, improving both training stability and material selection consistency. 
Finally, we have designed a new synthetic \dataset{DuMaS} dataset which comprises 800,000+ images of indoor and outdoor scenes.
It is over 16$\times$ larger than the recent Materialistic dataset~\cite{sharma2023materialistic} and includes annotations at two selection levels: texture and subtexture. 
The texture level is the same as in the Materialistic dataset and targets the selection of surfaces which belong to the same texture -- e.g., on a chess board, black and white squares would be selected together. Our new subtexture level adds a finer-grained option to enable \new{the selection of parts of textures with similar appearance, grouping together individual texture components} -- in the chess board example, black and white squares would be selected separately (see also \cref{fig:teaser}). This two-level approach also improves selection quality and consistency.

We evaluate the quality and consistency of our model both qualitatively and quantitatively, under different scenarios: varying the selected pixel, the image's field of view, or its lighting. 
We ablate our multi-resolution processing and training schemes as well as the impact of our dataset on selection accuracy. 
Finally, we compare our method to two state-of-the-art methods, namely Materialistic~\cite{sharma2023materialistic} and the Segment Anything Model 2 (SAM2) fine-tuned for materials~\cite{ravi2024sam,fischer2024sama}, showing significant improvement over both.

In summary, we propose a material selection model with improved accuracy and support for both texture- and subtexture-level selection in images thanks to the following contributions:
\begin{itemize}
    \item \new{A material selection architecture that offers two-level control of the selection granularity;}
    \item
        An improved training scheme for multi-pixel sampling and multi-resolution processing;
    \item
        A large synthetic dataset comprising annotations on both texture and subtexture levels. 
\end{itemize}
We will release our model and test code upon publication, as well as a significant subset of our dataset.

%%%%%%%%%%%%%%%%%%%%%%%%%%%%%%%%%%%%%%%%%%%%%%%%%%%%%%%%%%%%
\section{Related work}
%%%%%%%%%%%%%%%%%%%%%%%%%%%%%%%%%%%%%%%%%%%%%%%%%%%%%%%%%%%%

%%%%%%%%%%%%%%%%%%%%%%%%%%%%%%
\subsection{Object segmentation and selection}
%%%%%%%%%%%%%%%%%%%%%%%%%%%%%%

Recent research in object segmentation and selection has significantly advanced both 2D and 3D image understanding. 
In 2D images and videos, models such as SAM and SAM2 \cite{ravi2024sam, kirillov2023segment} enable selection/segmentation (and tracking) of objects. 
In 3D representations, methods for radiance fields \cite{kim2024garfield, bhalgat2023contrastive, fu2022panoptic}, point clouds \cite{tchapmi2017segcloud} and inter-surface mappings \cite{morreale2024neural} leverage geometric or multi-view cues to achieve segmentation and selection in a consistent manner.

These techniques, however, operate on an \emph{object} level, targeting the selection of distinct objects in an image.
In contrast, we target fine-grained material selection, where our goal is to identify regions sharing the clicked pixel's material, irrespective of the object(s) onto which the material is applied: a single object can contain multiple materials and a single material can appear on multiple objects.

%%%%%%%%%%%%%%%%%%%%%%%%%%%%%%
\subsection{Material segmentation and selection}
%%%%%%%%%%%%%%%%%%%%%%%%%%%%%%

Material selection from 2D images is a non-trivial problem, since a surface's perceived appearance can be influenced by numerous factors beyond its reflectance properties, such as its geometry \cite{boyaci2003effect}, illumination \cite{fleming2003real}, or the surrounding surfaces and object identity \cite{sharan2009perception, sharan2014accuracy}. 
Early material selection algorithms were built around low-level features such as color- and texture descriptors \cite{belongie1998color, haralick1973textural} or hand-crafted filters and heuristics \cite{leung2001representing, malpica2003multichannel}. 
Additionally, selection is simplified when the image can be decomposed into (potentially disjoint) regions of varying reflectance properties, as shown by \citet{lensch2003image} or, more recently for radiance fields, by \citet{verbin2022ref}.
However, this often requires additional information such as multi-view images or specialized capture hardware \cite{xue2020differential}, and single-image decomposition into physical components remains challenging~\cite{zeng2024rgb, zhu2022irisformer, kocsis2024iid}.

Leveraging deep networks' strong classification capabilities, methods targeted material semantic classification~\cite{bell2015material, cimpoi2014deep, sumon2022multi}. Others proposed to train material perceptual similarity metrics for classification of complete photographs containing materials ~\cite{sharan2013recognizing, lagunas2019similarity}.

Closest to our method is Materialistic~\cite{sharma2023materialistic}, a material selection method leveraging large vision models' features~\cite{caron2021dino}.
While we also target material selection, we differ from this work in multiple ways: we improve feature processing through a multi-resolution approach, preserving more signal throughout the pipeline and hence improving selection accuracy and consistency.  
Further, we extend their proposed definition of pixels with similar materials (i.e., pixels belonging to the same texture) by adding a finer-grained selection level we call subtexture. 
This enables the selection of texture sub-elements which share the same appearance. 
Additionally, despite the existence of %material- and semantic  
several datasets~\cite{murmann2019dataset, sharan2014accuracy, wang20164d, bell2013opensurfaces, upchurch2022dense, deschaintre2018single, vecchio2024matsynth, eppel2024learning} for semantic material classification or material selection, they typically provide coarse semantic annotations. 
For material selection, \citet{sharma2023materialistic} used 50,000 synthetic renderings with ground-truth annotations.
However, their dataset contains few, strongly textured materials with only texture-level annotations. 
In contrast, we design a significantly larger (800,000+ images) synthetic dataset containing many spatially varying textures with annotations both at the texture and subtexture levels.

%%%%%%%%%%%%%%%%%%%%%%%%%%%%%%
\subsection{Image encoders}
%%%%%%%%%%%%%%%%%%%%%%%%%%%%%%

Most modern object and material selection methods utilize the features of big vision models, ViTs or masked auto-encoders (MAEs) pre-trained on large collections of natural images, as backbones. 
ViTs like DINO and DINOv2 \cite{caron2021dino, oquab2024dinov2} learn priors encoded in these images (e.g., the appearance of shadows and reflections), making them well-suited for generalization to vision tasks such as object or material selection.
A defining feature of both ViTs and MAEs is the use of self-attention \cite{vaswani2017attention}, enabling global information sharing across the image.

However, attention is a memory-intensive operation, limiting both architectures in terms of input patch size and resolution. 
The recently introduced Hiera architecture \cite{ryali2023hiera, bolya2023hieradet} mitigates this via hierarchical feature extraction and smaller kernels, leading to sharper feature boundaries. Scaling-on-scales (S$^2$, \cite{shi2024we}), proposes to upscale the input to different resolutions and concatenate the resulting features, obtaining comparable (and occasionally superior) performance to larger vision models. 

In this work, we also leverage the features of big vision models, evaluate various options~\cite{caron2021dino, oquab2024dinov2, ryali2023hiera} and adjust the scaling proposed by S$^2$ to fit our selection context, significantly improving selection quality and consistency by preserving more of the input image's available information.

%%%%%%%%%%%%%%%%%%%%%%%%%%%%%%%%%%%%%%%%%%%%%%%%%%%%%%%%%%%%
\section{Method}
\label{sec:method}
%%%%%%%%%%%%%%%%%%%%%%%%%%%%%%%%%%%%%%%%%%%%%%%%%%%%%%%%%%%%

\begin{figure*}[t]
    \centering
    \includegraphics{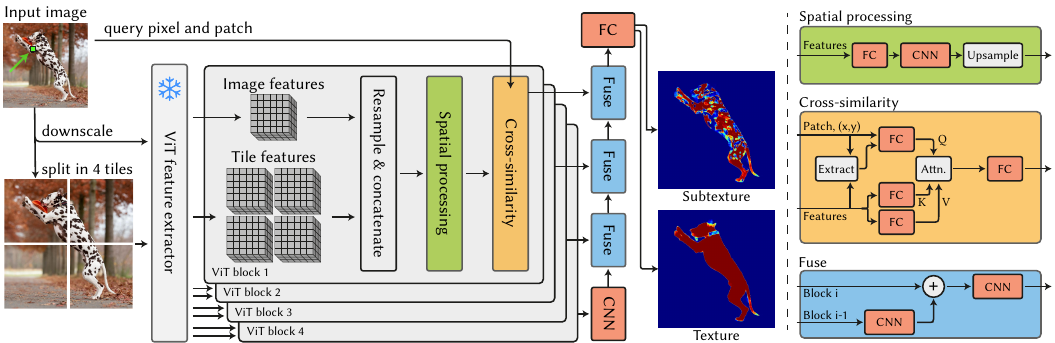}
    \vspace{-0.1cm}
    \caption{
        \textbf{Model architecture.} 
        Our model extracts ViT features at different resolutions, for the full image and for each separate tile, leading to a multi-resolution feature encoding. 
        The subsequent spatial processing layers upscale the features before the cross-similarity computes the attention with respect to the clicked image pixel and patch.
        We exploit the information encoded at different depths by repeating this process across four ViT levels, before fusing the output with a residual CNN and feeding it to our two-level selection head, producing selections at both subtexture and texture level. 
        The ViT is frozen, the red blocks are trained. 
        FC is short for fully-connected network.
    }
    \label{fig:overview2}
\end{figure*}

We design a new state-of-the-art model for material selection in images that addresses the precision and robustness limitations of prior approaches, and build a large synthetic dataset with material annotations at both texture and subtexture levels.

Our model takes as input an image and a query pixel, and outputs per-pixel material similarity to this query at both texture and subtexture levels, as demonstrated in \cref{fig:teaser}. This similarity can then be thresholded into a binary selection mask. We build our model on the architecture of Materialistic~\cite{sharma2023materialistic}, to which we make three key modifications, described in detail in~\cref{sec:model}: (1) multi-resolution processing and feature aggregation to improve precision, (2) two-level representation to allow for texture- and subtexture-level selection, and (3) multiple query sampling during training to improve robustness. 

An overview of our pipeline is shown in~\cref{fig:overview2}. 
We first extract features from the input image with a ViT encoder at different resolutions. Our model can be used with different encoders, and we evaluate alternatives in \cref{sec:results}; for convenience, unless explicitly stated, our explanations and results in the rest of the paper use DINOv2. 
The extracted features are then processed through our proposed multi-resolution aggregation scheme, and fed into a material selection head akin to that of Materialistic, modified for two-level selection. 
This head takes the aggregated features, and computes the cross-similarity between the features and the image patch centered around the user-provided query pixel, yielding query-conditioned features. 
Both in Materialistic and in our work, the feature extraction by the encoder -- and therefore the subsequent processing -- is done from four different transformer blocks. After the cross-similarity feature weighting layer, the query-conditioned features at different blocks are combined and bilinearly upsampled via convolutional layers. Finally, a channel-wise sigmoid is applied to yield the final per-pixel similarity. 
The model is trained minimizing a binary cross-entropy (BCE) loss on our DuMaS dataset (\cref{sec:dataset}).

%%%%%%%%%%%%%%%%%%%%%%%%%%%%%%%%%%%%%%%%%%%%%%%%%%%%%%%%%%%%
\subsection{Two-Level Multi-Resolution Material Selection}
\label{sec:model}
%%%%%%%%%%%%%%%%%%%%%%%%%%%%%%%%%%%%%%%%%%%%%%%%%%%%%%%%%%%%

We now describe here the three key components that enable our model to perform fine-grained, spatially varying material selection (\S\ref{subsubsec:multires} to \S\ref{subsubsec:multisampling}). Prior to that (\S\ref{subsubsec:encoder}), we discuss our feature extractor of choice and its advantages over previous alternatives.

%%%%%%%%%%%%%%%%%%%%%%%%%%%%%%
\subsubsection{Image encoder}
\label{subsubsec:encoder}
%%%%%%%%%%%%%%%%%%%%%%%%%%%%%%
Different encoders can be used to extract features from the input image. 
In this work, we rely on DINOv2~\cite{oquab2024dinov2}, a pre-trained self-supervised ViT that builds upon the foundation of DINO~\cite{caron2021dino}, incorporating improvements in architecture and training strategies. 
In particular, it utilizes a larger and more diverse dataset, better augmentations, and a more refined distillation process to enhance feature extraction capabilities with richer representations. 
DINOv2 also adopts a larger patch size ($ps = 14$, almost twice as large as DINO), which captures higher contextual information but results in lower-resolution tensors at $\sfrac{1}{14}$ of the input image resolution.
In our experiments, using DINOv2 as an encoder yields more discriminative features, particularly for disentangling appearance from lighting variations, and more confident selection predictions compared to using the original DINO.
Its bigger patch size slightly degrades performance on edges and small details on cluttered scenes when using a single resolution approach. 
However, this is mitigated by our multi-resolution approach, where a bigger patch size does not hamper performance. 
In the following, we refer to DINOv2 as the pre-trained embeddings from variant ViT-B/14 with $ps = 14$ and feature dimension $d=768$, our best performing configuration. Following previous work~\cite{sharma2023materialistic}, we extract four intermediate features (both local and global tokens) from transformer blocks at indices 2, 5, 8, and 11.

%%%%%%%%%%%%%%%%%%%%%%%%%%%%%%
\subsubsection{Multi-resolution processing and feature aggregation}
\label{subsubsec:multires}
%%%%%%%%%%%%%%%%%%%%%%%%%%%%%%

Despite the impressive performance of ViTs as feature extractors, their tokenization significantly reduces the resolution of the input images, degrading precision on edges and thin structures.
Therefore, we propose to extract features from regions of the input image at multiple resolutions, improving the sharpness and robustness of our material selection results.
Given an input image $I$, we construct a pyramid of $n$ resolutions $\{I_{1}, I_{2}\}$, where $I_{1}$ represents the image at the input resolution $r \times r$ of the image encoder ($r = 518$ for DINOv2), and $I_{i}$ represents $2^{i-1}$ higher-resolution versions, ensuring that their resolution remains divisible by the ViT patch size $ps$ (to maintain alignment between the feature maps across resolutions).
Each $I_{i}$ is split into $2^i$ non-overlapping tiles, 
each of them with the resolution of $I_{1}$, $r \times r$. \new{In practice, our $1024 \times 1024$ training images are downsampled to the image encoder's native resolution for $I_{1}$, while $I_{2}$ makes use of the full image resolution, slightly resized and split into four tiles.}
These tiles are fed into the image encoder, which, as explained in \S\ref{subsubsec:encoder}, extracts features at four different scales (from four transformer blocks, $j=1,...,4$). This yields feature maps $F_{ij}$, extracted at image resolution $i$ and block $j$. In our experiments, we evaluated two and three resolution levels and found that three levels did not provide tangible benefits given our training data resolution, while requiring significantly more compute for the additional tiles' features. We therefore use two levels $n=2$, effectively achieving twice higher resolution than the single resolution approach. 

We then introduce a feature aggregation module to integrate information across resolutions (see~\cref{fig:overview2}).
We first resample all feature maps $F_{ij}$ to match the target resolution (with different target resolutions for each block $j$) and concatenate the features along the channel dimension\new{, making sure to preserve the original image layout when re-arranging the tiles}:
\begin{equation}
    F_{\text{agg},j} = \text{Concat}(\{\text{Resample}(F_{ij})\}_{i=1}^{n}),
\end{equation}
where $\text{Resample}(\cdot)$ denotes bilinear upsampling and area downsampling to match the target resolution, and $\text{Concat}(\cdot)$ represents concatenation along the channel axis. 

\begin{figure}
    \centering
    \includegraphics{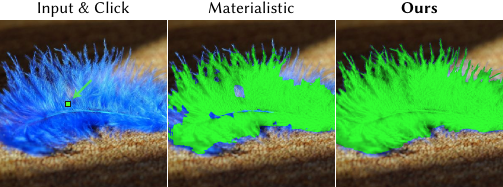}
    \caption{\textbf{Fine-grained material selection.} Our multi-resolution aggregation allows to recover thin structures, overcoming a limitation acknowledged in previous work~\cite{sharma2023materialistic}.
    }
    \label{fig:multires}
\end{figure}

Previous related work on multi-resolution aggregation~\cite{shi2024we} always downsamples the higher-resolution features before concatenating them. 
In contrast, we adapt our up- or down- sampling strategy to the various target feature resolutions per block, where we make sure to preserve the highest-resolution spatial details.
For aggregation, concatenation outperforms averaging as it preserves more information by retaining both fine-grained and broader contextual features that are later processed by the material selection head. We illustrate the benefits of this module in \cref{fig:multires},
where our multi-resolution processing allows to sharply select the thin structures of the feather.

%
%%%%%%%%%%%%%%%%%%%%%%%%%%%%%%
\subsubsection{Two-level representation}
\label{subsubsec:twolevel}
%%%%%%%%%%%%%%%%%%%%%%%%%%%%%%

Our model outputs material similarity representations at two levels, texture-level and subtexture-level, in the form of two different similarity maps. 
To achieve this, we modify the output per-pixel similarity score of our model to have two output channels, each of them using a sigmoid activation function. We train both channels jointly, minimizing the BCE loss on our \dataset{DuMaS} dataset.

%%%%%%%%%%%%%%%%%%%%%%%%%%%%%%
\subsubsection{Multi-query sampling during training}
\label{subsubsec:multisampling}
%%%%%%%%%%%%%%%%%%%%%%%%%%%%%%

To improve training stability and robustness, we sample multiple query pixels per image during training, covering diverse materials, which has proven helpful in the object selection context in previous work~\cite{ravi2024sam}. 
This strategy is similar in spirit to using a higher effective batch size, re-using the image encoder computation and making the optimization more stable.
We show in \cref{sec:results} how the multiple query sampling benefits the robustness and accuracy of our material selection results.

%%%%%%%%%%%%%%%%%%%%%%%%%%%%%%%%%%%%%%%%%%%%%%%%%%%%%%%%%%%%
\subsection{\dataset{DuMaS} training dataset}
\label{sec:dataset}
%%%%%%%%%%%%%%%%%%%%%%%%%%%%%%%%%%%%%%%%%%%%%%%%%%%%%%%%%%%%

\begin{figure*}
    \centering
    \vspace{-1mm}
    \includegraphics{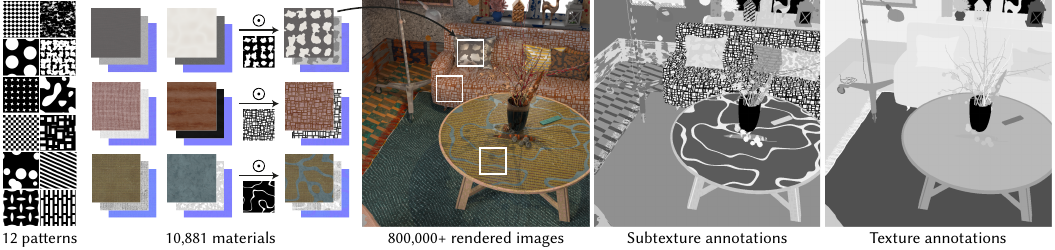}
    \vspace{-3mm}
    \caption{
        \textbf{\dataset{DuMaS} training dataset creation.} 
        Starting from twelve base noise patterns from Substance Designer, we vary their parameters to create 1,571 binary masks which we use to randomly combine 3,026 stationary reflectance maps to generate 10,881 materials (see text for more details). 
        We assign these material maps to objects of our 132 scenes under different combinations, and render videos by varying camera viewpoints, resulting in 816,415 individual images.
        For each image, our dataset includes dense annotations at both subtexture and texture level; annotation IDs are mapped to gray levels for visualization.
    }
    \label{fig:dataset-creation}
\end{figure*}

Most existing material datasets with dense image-space annotations \cite{bell2013opensurfaces,murmann2019dataset,upchurch2022dense} contain semantic annotations of material classes, such as ``wood'' or ``metal'', but lack fine-grained annotations of variations within a class. The dataset of Materialistic~\cite{sharma2023materialistic} does contain fine-grained per-pixel material annotations for 50,000 synthetic images; however, such annotations do not distinguish between different individual texture components within a single texture (e.g., black and white squares on a checkered pattern). Such distinction is necessary to enable our two-level material selection.
Therefore, we render a new large-scale synthetic dataset of over 800,000 images including fine-grained material annotations at both texture- and subtexture-level, named \dataset{DuMaS} (Dual-level Material Selection) dataset.

We do so by rendering 132 different synthetic scenes (119 indoor, 13 outdoor) from the Evermotion collection~\cite{evermotion} with random materials alongside texture and subtexture annotations.
We first create the material maps which will be applied to objects in different scenes. We select a set of 3,026 stationary\footnote{A class of materials consisting of plain colors, or structures that (randomly) repeat over the surface~\cite{aittala2015two}.} source materials from the Adobe 3D Assets library, constituting our initial set of \emph{reflectance} maps. 
We then create 1,571 different binary masks by randomly sampling generative parameters from twelve noise patterns in Substance Designer. 
Combining a binary mask (used as alpha channel) with a pair of reflectance maps leads to a new \emph{texture} map (see ~\cref{fig:dataset-creation}, left). In particular, for each binary mask we sample five pairs of reflectance maps without repetition, for a total of 7,855 different texture maps. Together, the reflectance and texture maps yield 10,881 unique material maps. Each reflectance and each texture map are assigned a unique ID, which is then used for image annotation at texture and subtexture levels. 

For each scene, we create five different material assignments by randomly sampling from our set of material maps, for a total of 660 different scene configurations.
We maintain the original relationships between objects, so that objects (or parts of them) with the same material in the original scene will also share the same material after our assignment. Transparent and emissive materials in the original scenes are left unmodified.
The final annotations, in the form of per-pixel material IDs at two levels, subtexture and texture, are as follows: if an object is assigned a material from the initial reflectance set, both levels will share the same ID.
If it is assigned a material from the texture set, the texture-level annotation will store its ID, and the subtexture level will store the ID of its assigned constituent reflectance map (see \cref{fig:dataset-creation}, right).

We render videos of up to one minute for every scene, following a camera trajectory that mimics a first-person exploration of the scene, at 30 fps.
Each frame is rendered at 1024$\times$1024 resolution with 256 samples per pixel using Blender Cycles 4.2.
Our full \dataset{DuMaS} dataset contains 816,415 frames (around 250 days of GPU rendering time).
Including videos instead of independent images allows us to use our dataset to fine-tune video selection models like SAM2.

%%%%%%%%%%%%%%%%%%%%%%%%%%%%%%%%%%%%%%%%%%%%%%%%%%%%%%%%%%%%
\subsection{Implementation Details}
%%%%%%%%%%%%%%%%%%%%%%%%%%%%%%%%%%%%%%%%%%%%%%%%%%%%%%%%%%%%

We train our model on our \dataset{DuMaS} dataset for 10 epochs, using the Adam optimizer with learning rate 1e-4 on four A100-40GB GPUs, using the DDPS distributed strategy and batch size 4 images per GPU.  
During training, we sample random crops at the ViT resolution and apply random exposure, saturation, and brightness augmentations. For our multi-query sampling, we uniformly sample 10 pixels within the crop.

%%%%%%%%%%%%%%%%%%%%%%%%%%%%%%%%%%%%%%%%%%%%%%%%%%%%%%%%%%%%
\section{Results}
\label{sec:results}
%%%%%%%%%%%%%%%%%%%%%%%%%%%%%%%%%%%%%%%%%%%%%%%%%%%%%%%%%%%%

In this section we present qualitative and quantitative evaluations, as well as a robustness analysis of our model with respect to user inputs, zoom levels, illumination, and sensitivity to the selection threshold. Finally, we ablate several aspects of our architecture \new{and show application examples for material editing at texture and subtexture levels}. \new{We will publicly release our evaluation framework's code and create a benchmark for both material selection quality and robustness, to facilitate future work}.

%%%%%%%%%%%%%%%%%%%%%%%%%%%%%%%%%%%%%%%%%%%%%%%%%%%%%%%%%%%%
\subsection{Real-World Test Datasets}
%%%%%%%%%%%%%%%%%%%%%%%%%%%%%%%%%%%%%%%%%%%%%%%%%%%%%%%%%%%%

We evaluate our method \new{mainly} on two test datasets that contain \emph{in-the-wild}, real-world images: (i) the \dataset{\materialisticds} dataset~\cite{sharma2023materialistic}, containing 50 images annotated at \textit{texture} level; and (ii) the \dataset{\ourrealds} dataset, our new, manually annotated test set containing 20 images with annotations at both \textit{subtexture} and \textit{texture} levels. This new dataset contains challenging real-life scenarios with strong lighting variations, indoor and outdoor instances, cluttered scenes with thin structures, and high-frequency appearances. We will release this new test set upon publication. 
In both datasets, we sample 10 query pixels per image for evaluation, generating 500 test cases for the \dataset{\materialisticds} dataset, and 200 test cases for our \dataset{\ourrealds} dataset.

\begin{figure*}[h]
 \centering
 \includegraphics[width=1\linewidth]{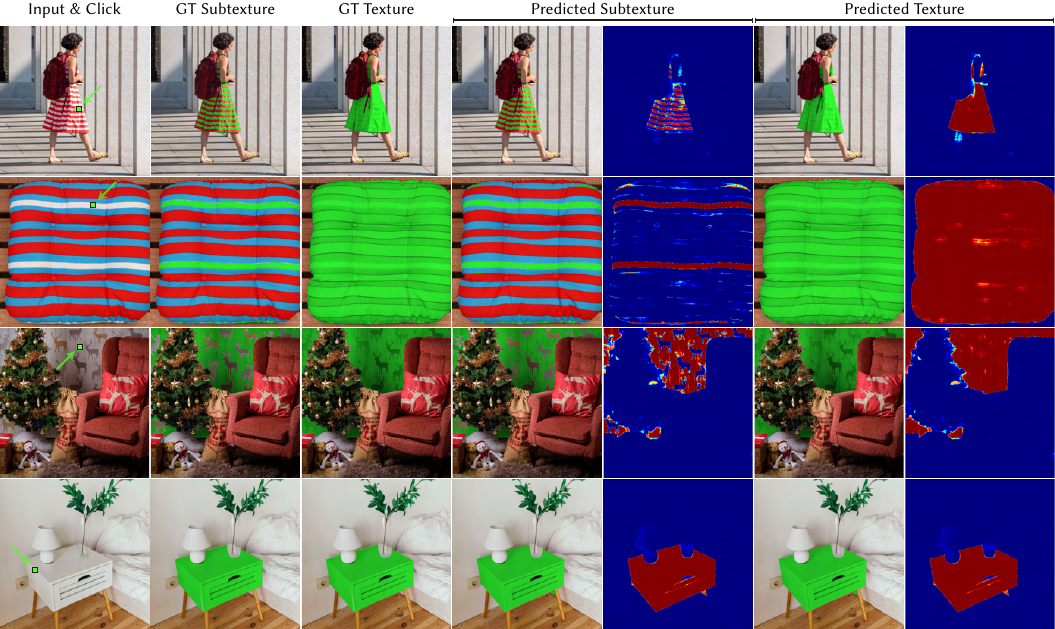}
 \caption{
 \textbf{Qualitative results}. We show results of our method (two-level selection) for images in our real-world test datasets. 
 For each example and level, we show the predicted, binarized selection masks as green image overlays and the similarity score before thresholding in false color (blue low, red high).
 Our method works well at both subtexture and texture levels even for cluttered scenes and patterns very different from the ones in the training set (third row, wallpaper), and with challenging examples 
 where objects in the scene share the same color (bottom row). 
 In non-spatially varying materials (bottom row) our model's predictions for both levels are consistently the same.
 }
 \label{fig:qualitative}
 \end{figure*}

\newpage

%%%%%%%%%%%%%%%%%%%%%%%%%%%%%%%%%%%%%%%%%%%%%%%%%%%%%%%%%%%%
\subsection{Evaluation}
%%%%%%%%%%%%%%%%%%%%%%%%%%%%%%%%%%%%%%%%%%%%%%%%%%%%%%%%%%%%
%%%%%%%%%%%%%%%%%%%%%%%%%%%%%%
\paragraph{Qualitative results}
%%%%%%%%%%%%%%%%%%%%%%%%%%%%%%
\Cref{fig:qualitative} presents the results of our two-level selection method in challenging scenarios. These images contain a diverse range of cases, demonstrating the robustness of our approach. 
Our method accurately selects textured areas and discriminates subtextures, as shown in rows one to three. 
The third row shows a highly cluttered scene, with several spatially-varying materials, in which our method successfully makes the right selection at both levels, despite the pattern being very different from the training ones. 
The last row showcases a challenging, mostly-white scene with varying reflectances, where our method correctly identifies the table material at both texture and subtexture levels.

%%%%%%%%%%%%%%%%%%%%%%%%%%%%%%
\paragraph{Comparisons} 
%%%%%%%%%%%%%%%%%%%%%%%%%%%%%%
%
\begin{figure*}
 \centering
 \includegraphics[width=\linewidth]{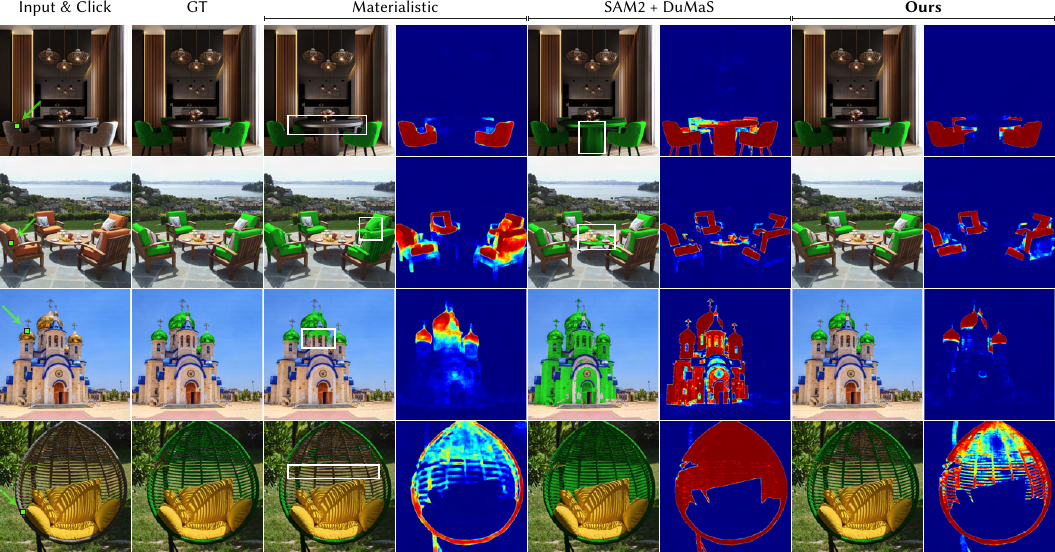}
 \caption{
 \textbf{Qualitative comparison.}
 Texture-level selection results for Materialistic, SAM2 fine-tuned on our \dataset{DuMaS} dataset, and our method.
 The white squares highlight areas where the methods struggle.
 Materialistic fails to select relevant areas, especially in the presence of small or thin structures (first and last rows), or to produce sharp, clear selections (second and third rows). 
 SAM2 has improved sharpness, but produces many false positives in areas of similar appearance.
 }
 \label{fig:qualitative_sota}
 \end{figure*}

We compare our model with two recent state-of-the-art methods: Materialistic \cite{sharma2023materialistic}, and SAM2~\cite{ravi2024sam}.
Since SAM2 was originally trained for \emph{object} selection, we fine-tune it with our \dataset{DuMaS} dataset to the material selection task. For completeness, we also report results of Materialistic trained on our \dataset{DuMaS} dataset, \new{as well as using DINOv2 as encoder}. 
For the SAM2 model evaluations we duplicate the first frame and use the output of the second frame~\cite{fischer2024sama}, which significantly improves its overall confidence and accuracy.

\Cref{fig:qualitative_sota} presents \textit{qualitative} results for texture-level selection on real images, from a single query pixel, comparing our method with Materialistic and SAM2 fine-tuned with our \dataset{DuMaS} dataset. Overall, our method is more accurate and results in fewer false positives.
Compared to Materialistic, it consistently produces sharper and cleaner predictions across all examples; this is probably due partly to our \dataset{DuMaS} dataset, but also to Materialistic's reliance on features at a single resolution, which constrains the precision of the output, and results in blurry edges and a diminished capacity to accurately select fine details.  
In contrast, the fine-tuned SAM2 model, using the Hiera encoder (with a smaller patch size and twice the input resolution), produces sharp edges and handles thin structures, but tends to over-select areas of the image of relatively similar appearance.

\Cref{tab:quantitative_comparisons} reports \textit{quantitative} results for three different metrics on both test datasets. 
The L1 metric measures the  pixel-wise difference between the predicted values and the ground truth mask; lower values indicate better agreement, which can be interpreted as higher prediction confidence. The Intersection over Union (IoU) measures the overlap between the ground truth mask and the predicted mask, and the F1 score is the harmonic mean of precision and recall, providing a single metric that balances both aspects of a model's accuracy.
IoU and F1 are computed by binarizing the outputs with a threshold, which we fix to 0.5.
Metrics for tasks outside a method's original training scope are shown in the table for reference, but marked in \textcolor{gray}{gray} (e.g., our subtexture level is not supported by previous works).
\Cref{tab:quantitative_comparisons} supports our qualitative comparisons; when trained on \dataset{DuMaS}, Materialistic slightly improves its performance, due to the larger and more diverse dataset. Still, our method consistently achieves the best performance for both subtexture and texture level selection. \new{Interestingly, when changing the image encoder of Materialistic to DINOv2 (second row), the overall quantitative performance of Materialistic is slightly degraded: despite the stronger feature representation capabilities of DINOv2, its larger patch size (14 vs. 8) yields lower resolution features, significantly impacting performance on edges. This highlights the benefits of our multi-resolution pipeline, compensating for this limitation.}

% using mean
\begin{table}[]
\caption{
    Mean results of various methods (rows) across different metrics (subcolumns) on two real-world test datasets (columns) . 
    For single-level methods trained on our \dataset{DuMaS} dataset (+\dataset{DuMaS}), we train two separate models (one per level) and report the result of the relevant one per column.
    Gray text indicates cases where the model is evaluated on a different task to the one it is trained for, and boldface marks the  best results.
    %\julia{TODO: combine the Two-Level columns (same dataset, two level labels)} \elena{It is confusing to see Materialistic both in the column (as the datset) and in the row (as a method). Maybe add Dataset at the end of Materialistic and Two-level, or add another single top row that reads Dataset because otherwise.} \julia{we can also change the "name" of the materialistic test dataset, it doesnt have any name} 
}
\setlength\tabcolsep{0.6mm}%
\resizebox{\columnwidth}{!}{%
\begin{tabular}{@{}llllllllll@{}}
\toprule
& \multicolumn{3}{c}{\dataset{\textbf{Materialistic Test}}} & \multicolumn{6}{c}{\dataset{\textbf{Two-Level Test}}} \\ 
& \multicolumn{3}{c}{\textbf{Texture Level}} & \multicolumn{3}{c}{\textbf{Subtexture Level}} & \multicolumn{3}{c}{\textbf{Texture Level}} \\ \cmidrule(l{2pt}r{2pt}){2-4} \cmidrule(l{2pt}r{2pt}){5-7} \cmidrule(l{2pt}r{2pt}){8-10}
& \hspace{2.9mm}\textbf{L1} $\downarrow$ & \textbf{IoU} $\uparrow$       & \textbf{F1} $\uparrow$        & \textbf{L1} $\downarrow$        & \textbf{IoU} $\uparrow$      & \textbf{F1} $\uparrow$      & \textbf{L1} $\downarrow$ & \textbf{IoU} $\uparrow$      & \textbf{F1} $\uparrow$      \\ \midrule
Materialistic & \hspace{2.9mm}0.057 & 0.858 & 0.906 &  \textcolor{gray}{0.144} &  \textcolor{gray}{0.513} &  \textcolor{gray}{0.629} &  0.112 &  0.657 &  0.749 \\
\new{Materialistic\,+\,DINOv2} &  \hspace{2.9mm}\new{0.069}  &  \new{0.838}  &  \new{0.890} &  \textcolor{gray}{0.202}  &  \textcolor{gray}{0.463} &  \textcolor{gray}{0.581} &  \new{0.135} &  \new{0.636} &  \new{0.728} \\
Materialistic\,+\,DuMaS &  \hspace{2.9mm}0.043  &  0.858  &  0.904 &  0.092  &  0.615 &  0.717 &  0.101 &  0.680 &  0.765 \\

% Mat. + DuMaS Reflect.   &  \textcolor{gray}{0.049}  & \textcolor{gray}{0.848}  &  \textcolor{gray}{0.902} &  0.092  &  0.615 &  0.717 &  \textcolor{gray}{0.146} &  \textcolor{gray}{0.539} &  \textcolor{gray}{0.655} \\

% Mat. + DuMaS Texture   &  0.043  &  0.858  &  0.904  &  \textcolor{gray}{0.141}  &  \textcolor{gray}{0.515} &  \textcolor{gray}{0.630} &  0.101 &  0.680 &  0.765 \\

\midrule

SAM2 *obj.sel.  &  \hspace{2.9mm}\textcolor{gray}{0.086}  &  \textcolor{gray}{0.633}  &  \textcolor{gray}{0.708}  &  \textcolor{gray}{0.121}  &  \textcolor{gray}{0.426} &  \textcolor{gray}{0.539} &  \textcolor{gray}{0.118} &  \textcolor{gray}{0.558} &  \textcolor{gray}{0.632} \\ 

SAM2 + DuMaS &  \hspace{2.9mm}0.060  &  0.784  &  0.847 &  0.103  &  0.576 &  0.681  &  0.071 &  0.730 &  0.799 \\

% SAM2 + DuMaS Reflect. &  \textcolor{gray}{0.072}  &  \textcolor{gray}{0.748} &  \textcolor{gray}{0.820}  &  0.103  &  0.576 &  0.681 &  \textcolor{gray}{0.115} &  \textcolor{gray}{0.627} &  \textcolor{gray}{0.713} \\

% SAM2 + DuMaS Texture &  0.060  &  0.784  &  0.847  &  \textcolor{gray}{0.123}  &  \textcolor{gray}{0.561} &  \textcolor{gray}{0.672} &  0.071 &  0.730 &  0.799 \\

\midrule

\textbf{Ours} &  \hspace{2.9mm}\textbf{0.030}  &  \textbf{0.896}  &  \textbf{0.935}  &  \textbf{0.071}  &  \textbf{0.673} &  \textbf{0.766} &  \textbf{0.069} &  \textbf{0.750} &  \textbf{0.823}  \\ 
\bottomrule
\end{tabular}%
}
\label{tab:quantitative_comparisons}
\end{table}
% %

%%% Robustness Table
%
\begin{table}[]
\caption{
Consistency results of various methods (rows) across different query pixels, zoom levels, and illuminations (subcolumns) on three real-world test datasets (columns). For single-level methods trained on our \dataset{DuMaS} dataset (+\dataset{DuMaS}), we train two separate models (one per level) and report the result of the relevant one per column.
All are mean Hamming distances (lower is better). Note that only consistency is measured here, without assessing accuracy. 
%Gray text indicates cases where the model is evaluated on a different task to the one it is trained for, and boldface marks the  best results.
}
\setlength\tabcolsep{0.2mm}%
\resizebox{\columnwidth}{!}{%
\begin{tabular}{@{}lccccccc@{}}
\toprule
& \multicolumn{2}{c}{\dataset{\textbf{Materialistic Test}}} & \multicolumn{4}{c}{\dataset{\textbf{Two-Level Test}}} & \dataset{\textbf{Multi-Illumination}}  \\ 
& \multicolumn{2}{c}{\textbf{Texture Level}} & \multicolumn{2}{c}{\textbf{Subtexture Level}} & \multicolumn{2}{c}{\textbf{Texture Level}} & \textbf{Texture Level}\\ 
\cmidrule(l{2pt}r{2pt}){2-3} \cmidrule(l{2pt}r{2pt}){4-5} \cmidrule(l{2pt}r{2pt}){6-7} \cmidrule(l{2pt}r{2pt}){8-8}
 & \hspace{3.9mm}\textbf{Pixel} $\downarrow$       & \textbf{Zoom} $\downarrow$       & \textbf{Pixel} $\downarrow$       & \textbf{Zoom} $\downarrow$   & \textbf{Pixel} $\downarrow$       & \textbf{Zoom} $\downarrow$   & \textbf{Illumination} $\downarrow$    \\ \midrule
Materialistic  & \hspace{3.9mm}0.041 & 0.121 & \textcolor{gray}{0.082} &  \textcolor{gray}{0.200} &  0.082 &  0.200 &  0.071 \\

Materialistic\,+\,DuMaS   &  \hspace{3.9mm}0.028  &  0.087  & 0.077 &  0.222  &  0.059 &  0.161 & 0.065  \\

% Mat. + DuMaS Reflect.  & \textcolor{gray}{0.039}  &  \textcolor{gray}{0.122} & 0.077 &  0.222 &  \textcolor{gray}{0.077} &  \textcolor{gray}{0.222} &  - \\

% Mat. + DuMaS Texture  &  0.028  &  0.087 & \textcolor{gray}{0.059} &  \textcolor{gray}{0.161} &  0.059 &  0.161 &  - \\

%\midrule

SAM2\,+\,DuMaS  &  \hspace{3.9mm}0.025  &  0.080 &  0.053  &  \textbf{0.165}  &  0.059 &  \textbf{0.111} &  0.039 \\

% SAM2 *obj.sel.   &  \textcolor{gray}{0.002}  &  \textcolor{gray}{0.040}  & \textcolor{gray}{0.004} &  \textcolor{gray}{0.048} &  \textcolor{gray}{0.004} &  \textcolor{gray}{0.048} &  - \\

% SAM2 + DuMaS Reflect. & \textcolor{gray}{0.026} &  \textcolor{gray}{0.102} &  0.053  &  0.165 &  \textcolor{gray}{0.053} &  \textcolor{gray}{0.165} &  - \\

% SAM2 + DuMaS Texture   &  0.025  &  0.080  &  \textcolor{gray}{0.059}  &  \textcolor{gray}{0.111} &  0.059 &  0.111 &  - \\

\midrule

\textbf{Ours}  &  \hspace{3.9mm}\textbf{0.024}  &  \textbf{0.071}  &  \textbf{0.052} &  0.166 &  \textbf{0.041} &  0.112 &  \textbf{0.034} \\
\bottomrule
\end{tabular}%
}
\label{tab:consistency}
\end{table}
%

%%%%%%%%%%%%%%%%%%%%%%%%%%%%%%%%%%%%%%%%%%%%%%%%%%%%%%%%%%%%
\subsection{Robustness Evaluation}
%%%%%%%%%%%%%%%%%%%%%%%%%%%%%%%%%%%%%%%%%%%%%%%%%%%%%%%%%%%%
To evaluate the robustness of our method, we assess prediction \emph{consistency} across query pixels, zoom levels, and different illuminations.
Moreover, we evaluate how robust the predictions are to different thresholds, which we call prediction \emph{confidence}.

\begin{figure*}
 % \vspace{-10mm}
 \centering
 \includegraphics[width=\textwidth]{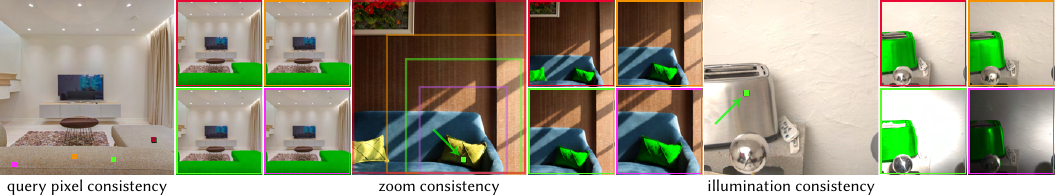}
  \caption{\textbf{Robustness.} Robustness evaluation of our method with respect to the clicked pixel (left subset), the image crop with increasing zoom levels (middle subset), and illumination changes (right subset).
  The images are challenging due to similar albedo (left), strong shading variations on the cushions (middle) and specular highlights on the toaster (right).
  }
 \label{fig:robustness}
 \end{figure*}

To test \emph{pixel consistency}, we randomly sample up to three materials per image, and five query pixels per material:
Ideally, the selection predictions for all query pixels belonging to the same material would be identical. 
With regard to \emph{zoom consistency},  we create five different crops with increasingly large zoom levels centered on each query pixel, and evaluate to what extent the same query pixel results in the same selection in all cases.
For \emph{illumination consistency}, we use the \dataset{Multi-Illumination}  dataset~\cite{murmann19ill}, which includes thirty in-the-wild scenes captured under twenty-five different illuminations. We sample up to three materials per scene and measure consistency of the selections across illuminations.

For all consistency evaluations, we compute the average pairwise Hamming distance (after binarizing the masks with a threshold of $0.5$), with lower Hamming distances indicating higher consistency.

As shown in \cref{tab:consistency}, our method demonstrates significantly higher consistency compared to Materialistic, achieving approximately $1.8\times$ lower Hamming distance. When Materialistic is trained on our \dataset{DuMaS} dataset, its consistency improves substantially, particularly across query pixels. This improvement underscores the benefits of our larger-scale dataset, which features a more diverse range of materials. Our architecture and training method further enhance consistency, especially noticeable on the more challenging \dataset{\ourrealds} and \dataset{Multi-Illumination} datasets.
Comparing to SAM2 fine-tuned on our \dataset{DuMaS} dataset, our method shows a slight improvement in consistency; moreover, as showed in \cref{tab:quantitative_comparisons}, our results are more accurate.

\cref{fig:robustness} shows qualitative examples for pixel, zoom and illumination consistency. In difficult scenarios, such as a sofa surrounded by surfaces of similar color (left), or a cushion with sharp shadows (middle), our predictions remain consistent and highly confident, clearly outperforming previous work. Regarding illumination, our predictions remain reasonably consistent even under very strong changes, as shown in \cref{fig:robustness} (right).

Finally, we evaluate the robustness of our method in terms of prediction \emph{confidence} by measuring the sensitivity of the selection result (the final binary mask) to the threshold over the similarity score, where our approach outperforms prior works.

%%%%%%%%%%%%%%%%%%%%%%%%%%%%%%%%%%%%%%%%%%%%%%%%%%%%%%%%%%%%
\subsection{Ablations}
\label{sec:ablations}
%%%%%%%%%%%%%%%%%%%%%%%%%%%%%%%%%%%%%%%%%%%%%%%%%%%%%%%%%%%%

% using mean
\begin{table}[]
\caption{Mean results of ablations of our method (rows) across different metrics (subcolumns) on the two test real-world evaluation datasets \new{and a challenging subset} (columns). For the Single Level ablation, we train two separate models (one per level) and report the result of the relevant one per column.
}
\setlength\tabcolsep{0.5mm}%
\resizebox{\columnwidth}{!}{%
\begin{tabular}{@{}lcccccccccc@{}}
\toprule
& \multicolumn{2}{c}{\dataset{\textbf{Materialistic Test}}} & \multicolumn{4}{c}{\dataset{\textbf{Two-Level Test}}} & \multicolumn{4}{c}{\dataset{\new{\textbf{Challenging Subset}}}} \\ 
& \multicolumn{2}{c}{\textbf{Texture}} & \multicolumn{2}{c}{\textbf{Subtexture}} & \multicolumn{2}{c}{\textbf{Texture}} & \multicolumn{2}{c}{\new{\textbf{Subtexture}}} & \multicolumn{2}{c}{\new{\textbf{Texture}}} \\ \cmidrule(l{2pt}r{2pt}){2-3} \cmidrule(l{2pt}r{2pt}){4-5} \cmidrule(l{2pt}r{2pt}){6-7} \cmidrule(l{2pt}r{2pt}){8-9} \cmidrule(l{2pt}r{2pt}){10-11}
& \hspace{4.9mm}\textbf{L1} $\downarrow$ & \textbf{IoU} $\uparrow$        & \textbf{L1} $\downarrow$        & \textbf{IoU} $\uparrow$   & \textbf{L1} $\downarrow$ & \textbf{IoU} $\uparrow$  & \new{\textbf{L1} $\downarrow$}        & \new{\textbf{IoU} $\uparrow$}   & \new{\textbf{L1} $\downarrow$} & \new{\textbf{IoU} $\uparrow$}    \\ \midrule

\textbf{Ours}, DINO   &  \hspace{4.9mm}0.045  &  0.850  &  0.094  &  0.616  &  0.097 &  0.698 & \new{0.107} & \new{0.591} & \new{0.104} & \new{0.644} \\

\textbf{Ours}, Hiera   &  \hspace{4.9mm}0.046  &  0.838  &  0.093  &  0.593 &  0.080 &  0.716 & \new{0.112}  & \new{0.581}  & \new{0.099}  & \new{0.685} \\

\textbf{Ours}, w/o Multi-Res. &  \hspace{4.9mm}0.033  &  0.889  &  0.081  &  0.637 &  0.075 &  0.740 & \new{0.130}  & \new{0.512}  & \new{0.122}  & \new{0.585}  \\

\textbf{Ours}, w/o Multi-Sampl.   &  \hspace{4.9mm}0.036  &  0.893  &  0.083  &  0.622  &  0.088 &  0.680 & \new{0.133}  & \new{0.507}  & \new{0.143}  & \new{0.517}  \\ 

\textbf{Ours}, Single Level   &  \hspace{4.9mm}0.037  &  0.888  &  0.077  & 0.643 &  0.081 &  \textbf{0.750} & \new{0.106}  & \new{0.581}  & \new{0.099}  & \new{0.685}  \\

% Ours, single level  &  \textcolor{gray}{0.039}  &  \textcolor{gray}{0.860}  &  \textcolor{gray}{0.908}  &  0.077  & 0.643 &  0.742 & \textcolor{gray}{0.135} &  \textcolor{gray}{0.550} &  \textcolor{gray}{0.663} \\

% Ours, single level   &  0.037  &  0.888  &  0.929  &  \textcolor{gray}{0.136}  &  \textcolor{gray}{0.572} &  \textcolor{gray}{0.677} &  0.081 &  \textbf{0.750} &  \textbf{0.824} \\

\midrule

\textbf{Ours}, Full &   \hspace{4.9mm}\textbf{0.030}  &  \textbf{0.896}   &  \textbf{0.071}  &  \textbf{0.673} &  \textbf{0.069} &  \textbf{0.750}  & \new{\textbf{0.068}}  & \new{\textbf{0.694}}  & \new{\textbf{0.058}}  & \new{\textbf{0.763}} \\ 
\bottomrule
\end{tabular}%
}
\label{tab:quantitative_ablations}
\end{table}
% %

We next analyze the impact of our most critical design decisions.
All variants have been trained the same number of epochs with the full \dataset{DuMaS} dataset unless otherwise stated. Our full model includes the DINOv2 image encoder, multi-resolution feature aggregation (Multi-res., \cref{subsubsec:multires}), and multiple query sampling (Multi-sampl., \cref{subsubsec:multisampling}). 

\begin{figure*}
 \centering
 \includegraphics[width=\textwidth]{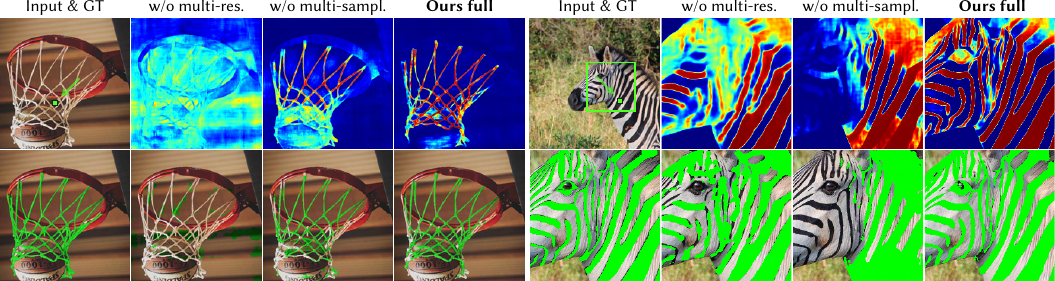}
  \caption{\textbf{Ablations.} We ablate parts of our method (multi-resolution and multi-sampling, respectively) \new{on two challenging examples containing thin structures (net and zebra stripes) and albedo entanglement (net has a similar albedo to the white parts of the basketball, in front of a patterned background).}
  }
 \label{fig:ablations}
 \end{figure*}

We first explore the effect of the image encoder (\cref{tab:quantitative_ablations}, first two rows). Replacing DINOv2 features with DINO or Hiera encoding produces less accurate selections and with noticeably less confidence (lower L1) in all datasets, as both DINO and Hiera seem to be more biased towards color and lighting than DINOv2.
Then, we show results removing our multi-resolution and multi-sampling approaches, both in \cref{tab:quantitative_ablations} and in \cref{fig:ablations}.
Our multi-resolution feature aggregation improves quality of the predictions (lower L1), which is particularly visible in the qualitative results, as it drastically improves performance when dealing with thin edges, like the basketball net. 
Our multi-sampling strategy, on the other hand, improves overall precision, in particular for texture-level selection. This may stem from computing gradients on the predicted selection for multiple materials in an image at a time, in every optimization step. Additionally, this multi-sampling significantly improves confidence by reducing the sensitivity to the selection threshold, which minimizes the need for manual adjustments.

We also assess in \cref{tab:quantitative_ablations} the effect of jointly training both subtexture and texture levels in one model (Ours Full) compared to training two separate models with a single output, one per level (Ours, Single Level). 
Notably, training with all data concurrently does not negatively impact performance, allowing to have a single model for both selection levels, and supporting our hypothesis that jointly estimating both outputs is beneficial for accuracy.

\new{Last, we further evaluate our ablations on a subset of 30 challenging test cases including fine structures, albedo entanglement, and strong light variations. The results (\cref{tab:quantitative_ablations}, right-most columns) show the clear benefits of our multi-resolution component, which significantly improves performance on thin structures, as well as our multi-sampling strategy, which improves robustness overall and  helps in scenarios with albedo entanglement and strong light variations (50\% lower L1 and 20\% higher IoU in \cref{tab:quantitative_ablations}). We hypothesize that by sampling multiple query pixels per image during training, it is more likely that these challenging cases (e.g., two pixels with same albedo from different materials, or two pixels from the same material in areas with strong lighting variation) appear in the same training batch, improving the gradient.}

\begin{figure}
    \centering
    \includegraphics[width=1\columnwidth]{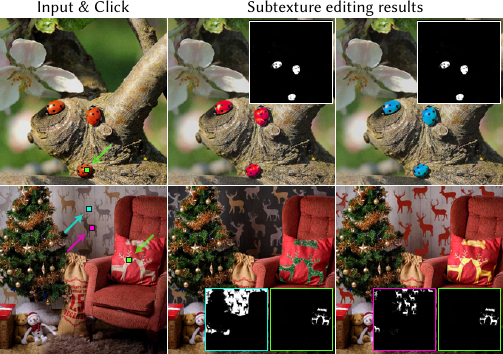}
    \caption{
        \new{\textbf{Editing.} We use our method's selection masks at subtexture level to perform fine-granular edits of the image's materials in Photoshop.}
    }
    \label{fig:editing}
\end{figure}
%

%%%%%%%%%%%%%%%%%%%%%%%%%%%%%%%%%%%%%%%%%%%%%%%%%%%%%%%%%%%%
\subsection{\new{Application: Material Editing}}
%%%%%%%%%%%%%%%%%%%%%%%%%%%%%%%%%%%%%%%%%%%%%%%%%%%%%%%%%%%%

\new{Selection of specific regions within an image is an exceedingly common and highly useful tool for a number of applications, among which editing is one of the most representative examples. We show practical applications of our selection for material editing tasks in image space, in \cref{fig:teaser} (bottom right) and \cref{fig:editing}. Our new level of granularity at the subtexture level, as well as our improved precision and accuracy, allow users to easily edit challenging scenarios that would otherwise require significant manual intervention, including individual texture components such as the flowers of the dress (\cref{fig:teaser}), or the deers on the cushion and wallpaper (\cref{fig:editing}).}

%%%%%%%%%%%%%%%%%%%%%%%%%%%%%%%%%%%%%%%%%%%%%%%%%%%%%%%%%%%%
\section{Discussion}
\begin{figure}
 \centering
    \includegraphics[width=1\columnwidth]{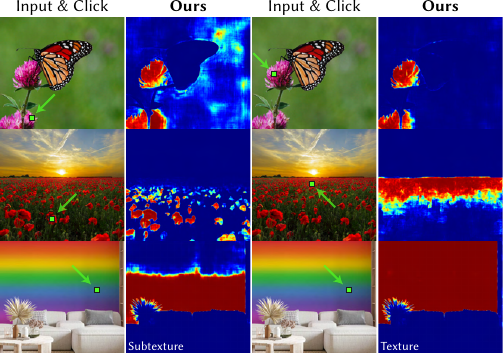}
    \caption{
        \textbf{Limitations.} Our method struggles with clicks on out-of-focus regions (top) and long-horizon imagery with changing frequencies (middle). \new{Textures without individual components (bottom) stretches our definition of subtexture.}}
    \label{fig:limitations}
\end{figure}

%%%%%%%%%%%%%%%%%%%%%%%%%%%%%%%%%%%%%%%%%%%%%%%%%%%%%%%%%%%%
\new{Several avenues of future work remain open.} Clicking on a pixel in an out-of-focus area of the image may lead to incorrect selections, as shown in the top row of \cref{fig:limitations}: as it can be seen, our model cannot accurately distinguish the flower petals and activates inaccurate regions in the background (left). The problem mostly goes away when clicking on an in-focus area of such flowers (right). Images with long horizon lines may also be problematic, if the image exhibits strongly varying frequencies due to perspective. In the middle row of \cref{fig:limitations} the first ranks of flowers are selected but the more distant ones are not (left), or viceversa (right). We believe these limitations could be addressed by adding more training data with explicit depth of field effects and outdoor large scenes, respectively. \new{Further, our definition of subtexture may not easily translate to continuously varying surfaces like the rainbow wall in the bottom row of \cref{fig:limitations}. Future work could explore a more continuous similarity definition to enable a gradient in the similarity score, following that of the color distance on the wall. More generally, while we propose an additional selection level, we do not claim to have decisively solved the inherent ambiguity of material selection tasks, for which different definitions of what ``the same material'' means may be needed, depending on the intended goal and downstream applications.}

\newpage
\section{Conclusion}
\new{We present a novel method for fine-grained material selection in images, which works both at texture and subtexture levels, and is more precise and robust than previous approaches. We evaluate various ViT backbones and propose a new training scheme and a large-scale dataset, significantly improving selection quality for fine structures and challenging scenarios with albedo entanglement and complex light variation.}

\begin{acks}
This work has been partially supported by grant PID2022-141539NB-I00, funded by MICIU/AEI/10.13039/501100011033 and by ERDF, EU. 
This work has also received funding from the Government of Aragon's Departamento de Educacion, Ciencia y Universidades through the Reference Research Group 
``Graphics and Imaging Lab'' (ref T34\_23R) and through the project ``HUMAN-VR: Development of a Computational Model for Virtual Reality Perception'' (PROY\_T25\_24).
Julia Guerrero-Viu developed part of this work during an Adobe internship, and was also partially supported by the FPU20/02340 predoctoral grant.
We thank the members of the Graphics and Imaging Lab, especially Nacho Moreno, Nestor Monzon, and Santiago Jimenez, for insightful discussions, help preparing the figures and final proofreading. 
% Nacho Moreno helped with figures, Nestor Monzon and Santiago Jimenez with proofreading
\end{acks}

\newpage
%%%%%%%%%%%%%%%%%%%%%%%%%%%%%%%%%%%%%%%%%%%%%%%%%%%%%%%%%%%%
% References
%%%%%%%%%%%%%%%%%%%%%%%%%%%%%%%%%%%%%%%%%%%%%%%%%%%%%%%%%%%%
\bibliographystyle{ACM-Reference-Format}
\bibliography{references}
\newpage

%%%%%%%%%%%%%%%%%%%%%%%%%%%%%%%%%%%%%%%%%%%%%%%%%%%%%%%%%%%%
\end{document}